\documentclass[final,1p,times,number]{elsarticle}
\usepackage{amsmath} %Environment aligned undefined. \begin{align}
\usepackage{booktabs} %表格环境
\usepackage{fontenc}
\usepackage{hyperref}
\usepackage{caption}
\usepackage{subcaption}
\newtheorem{lem}{Lemma}
\newtheorem{thm}{Theorem}
\newproof{pf}{Proof}

\newtheorem{remark}{Remark}
\newtheorem{assumption}{Assumption}
\usepackage{hyperref}
\usepackage{diagbox}
\usepackage{amssymb}
\usepackage{color}
\usepackage[utf8]{inputenc}
\usepackage{algorithm}
\usepackage{algpseudocode}
\usepackage{colortbl} 
\begin{document}
		
\begin{frontmatter}
			
			
			
\title{A spectral based goodness-of-fit test for stochastic block models}
			
			
\author{Qianyong Wu and Jiang Hu$^*$ }
\cortext[mycorrespondingauthor]{Corresponding author.}

\address{School of Mathematics $\&$ Statistics, Northeast Normal University, Changchun, China}

\begin{abstract}
Community detection is a fundamental problem in complex network data analysis. Though many methods have been proposed, most existing methods require the number of communities to be the known parameter, which is not in practice. In this paper, we propose a novel goodness-of-fit test for the stochastic block model. The test statistic is based on the linear spectral of the adjacency matrix.  Under the null hypothesis, we prove that the linear spectral statistic converges in distribution to $N(0,1)$. Some recent results in generalized Wigner matrices  are used to prove the main theorems. Numerical experiments and real world data examples illustrate that our proposed linear spectral statistic has good performance.
				
\end{abstract}
			
			
			
			
\end{frontmatter}
		
		
\section{Introduction}\label{intro}
Network data has been found in diverse areas,
such as social networks, gene regulatory networks, food webs, and many others \cite{jalanRandomMatrixAnalysis2007,jiCoauthorshipCitationNetworks2016,newmanNetworks2018,pontesBiclusteringExpressionData2015,westveldMixedEffectsModel2011}. One important topic in network analysis is detecting the community structure. 
The stochastic block model(SBM)  \cite{hollandStochasticBlockmodelsFirst1983,karrerStochasticBlockmodelsCommunity2011}  is one of the most popular and useful tools to model large networks with community structures. For an undirected network $G=(V,E)$, we give a brief introduction about how to generate the adjacency matrix by the SBM. We suppose all the $n$ nodes can be divided into $K$ disjoint communities as follows:
$$
V=V^{1}\cup V^{2}\cup \cdots \cup V^{K}.
$$
Denote $A$ as the $n\times n$ adjacency matrix of $G$. In the SBM  entries $A_{ij}(i > j)$  of the symmetric adjacency $A$ are independent
random variables satisfying
$$
P(A_{ij}=1)=1-P(A_{ij}=0)=B_{g_{i}g_{j}},
$$
where $g \in \left\{ 1,2,3 \cdots K \right\}^{n} $ is the label vector, $B = [B_{ab}]$ is a $K\times K$ symmetric matrix.  Here $B$ is a probability matrix that controls edge probabilities between communities. In this paper, we do not consider self-loops, as a result, we let the diagonal entries of the adjacency matrix be 0.
		
Community detection is recovering the label vector $g$ while giving a single observation of $A$. This problem has received considerable attention from different research areas, many methods have been proposed such as spectral clustering \cite{jinFastCommunityDetection2015,maSemisupervisedSpectralAlgorithms2018,leiConsistencySpectralClustering2015}, likelihood methods \cite{aminiPseudolikelihoodMethodsCommunity2013,bickelAsymptoticNormalityMaximum2013,newmanMixtureModelsExploratory2007} and modularity maximization \cite{newmanFindingCommunityStructure2006}, see  \cite{abbeCommunityDetectionStochastic2018} for a review. However, there exists a common flaw for most methods, which they assume the number of clusters is a known parameter which we do not. To solve this problem, Bickel et al. \cite{bickelHypothesisTestingAutomated2016} proposed to test $K = 1$ vs $K > 1$, under the null hypothesis, the largest eigenvalue test statistic that they proposed converges to  Tracy-Widom law with index one. Lei \cite{leiGoodnessoffitTestStochastic2016} extend this hypothesis test to a more general situation that is to test $K = K_{0}$ vs $K > K_{0}$. However, both the test statistics they proposed approach the limiting Tracy–Widom distribution slowly. In order to deal with the low convergence rate they adopt an additional bootstrap step. As a result it is time-consuming, especially for a very large network. Recently Dong et al. \cite{dongSpectralBasedHypothesis2020} proposed a novel test statistic to test whether the network is Erdős–Rényi graph, the test statistic is a linear spectral statistic of the adjacency matrix $A$.  Under the null hypothesis the proposed linear spectral statistic converges to $N(0,1)$, more importantly when the network size $n$ is small it still has good performance, so one can omit the bootstrap step. Inspired by Lei \cite{leiGoodnessoffitTestStochastic2016}, we extend Dong et al.'s \cite{dongSpectralBasedHypothesis2020} work to test $K = K_{0}$ vs $K > K_{0}$.  
		
We organize the rest of this paper as follows. The linear spectral statistic about the adjacency matrix is proposed in Section  \ref{MR}, and its asymptotic null distribution is also derived.  The sequential testing algorithm is put forward  to determine the number of communities. To illustrate the performance of the test statistic, some simulations and real world  examples are given in Section \ref{numexp} and \ref{realdata}. The technical proofs of the theorems are shown in Section \ref{pf}. We conclude this paper in section \ref{conclusion}.

\section{Main results}\label{MR}
\subsection{The linear spectral statistic}
Suppose we have a network of $n$ vertices and its adjacency matrix is represented mathematically by $A$. There is a question of whether we can fit this network by the stochastic block model with a hypothesis $K_{0}$ communities. Assume $K$ is the true cluster number, which is unknown in advance, this question can be solved in the following
hypothesis test framework:
\begin{align}\label{eq1} 
H_0:K=K_{0}  \quad vs \quad H_1:K\textgreater K_{0}.
\end{align}
To derive a goodness-of-fit statistic, we need to introduce some definitions and recent progress in random matrix theory (RMT).

Consider the $n\times n$ matrix $P$ given by 
$P_{ij}=B_{g_{i}g_{j}}$ we  center each $A_{ij}$ by subtracting $P_{ij}$ and  rescale it by dividing$ \sqrt{nP_{ij}(1-P_{ij})}$. 
Denote $\widetilde{A}$ be
\begin{align}\label{eq2} 
\widetilde{A}_{ij}=\left\{
\begin{array}{rcl}
\frac{A_{ij}-P_{ij}}{\sqrt{nP_{ij}(1-P_{ij})}}     &      & {i     \neq     j},\\
0   &      & {i=j}.\\
\end{array} \right.
\end{align}
Now we introduce our  test statistic
\begin{align}\label{eq3} 
T = \frac{1}{\sqrt{6}}trace(\widetilde{A}^{3}),
\end{align}
where $trace$ represents the  trace operator.
		
This processed matrix $\widetilde{A}$  is the so-called generalized Wigner matrix, satisfying $E(\widetilde{A}_{ij}) =0$ and $Var(\widetilde{A}_{ij}) =\frac{1}{n}$ for all $(i\neq j)$.
%The eigenvalues of $\widetilde{A}$ are denoted by $\lambda_1 \geq \lambda_2 \geq \cdots \geq \lambda_n$. 
The limiting spectral distribution(LSD) and linear spectral statistics(LSS) of Wigner matrix are very important topics in random matrix theory literature. Especially combining recent developments in \cite{wangGeneralizationCLTLinear2021} and  \cite{baiSpectralAnalysisLarge2010a}, we have the Theorem \ref{thm1}. We postpone the proof of Theorem \ref{thm1} to Section \ref{pf}.
\begin{thm}\label{thm1}
$\quad T = \frac{1}{\sqrt{6}}trace(\widetilde{A}^{3}) \stackrel{d}{\longrightarrow} N(0,1)$, where $ \stackrel{d}{\longrightarrow} $ means convergence in distribution.
\end{thm}
\begin{remark}
Theorem \ref{thm1} is a nontrivial generalization of Theorem 1 in  Dong et al. \cite{dongSpectralBasedHypothesis2020}.  Next, we outline the main differences between the two results. First, 
Dong et al. \cite{dongSpectralBasedHypothesis2020}  introduce new random variables $C_{ii}$ and let $A_{ii}=C_{ii}$, where $C_{ii}$ are i.i.d  random variables satisfying $P(C_{ii}=\frac{1}{\sqrt{n}})=P(C_{ii}=-\frac{1}{\sqrt{n}})=\frac{1}{2}$ for  $1\leq i\leq n$. In this paper, we do not require such unnecessary random variables. Second, the proof in  \cite{dongSpectralBasedHypothesis2020} needs the so-called homogeneity of the fourth moments: $E[\widetilde{A}_{ij}^4]=M<\infty$, for $1\leq i\neq j\leq n$. However,
the entries of $\widetilde{A}$ do not satisfy this condition when $K>1$. Thus, in this paper, we adopt the recent RMT results in  \cite{wangGeneralizationCLTLinear2021} to avoid the homogeneity of the fourth moments condition.
Most importantly, in order to find the number of communities,   Dong et al. need a multiple tests procedure. However, when multiple tests are conducted simultaneously, one has to resort to some corrections  to control the overall Type I error rate (e.g., Bonferroni correction).  It is known that the corrections may be conservative if there are a large number of tests \cite{johnsonAppliedMultivariateStatistical2007}. 
Through our Theorem \ref{thm1}, we can get an algorithm (Algorithm \ref{alg1} below) that does not require multiple tests.  
			
\end{remark}
		
In real network datasets we have no information about the true parameter $P$. As a result, one cannot  use
$T$ as the test statistic. In the following, we give an estimated  $\hat{T}$ by plugging in an estimated $\hat{P}$ and prove that the estimated test statistic $\hat{T}$ still holds asymptotic normality in Theorem \ref{thm2}.
		
Given a hypothesis $K=K_0$, let $\hat{g}$ be a strongly consistent estimated label vector by using some community detection methods such as spectral clustering \cite{vonluxburgTutorialSpectralClustering2007}. Define $\hat{n}_{k}= \#( \hat{V}_{k} )$, and $\hat{V}_{k}=\left\{i:1\leq i \leq n, \hat{g}(i)=k\right\},$ for $1\leq k \leq K_0$. One can get the estimated $B$ as following \cite{leiGoodnessoffitTestStochastic2016}:
		
$$
\hat{B}_{kl}=\left\{
\begin{array}{rcl}
\frac{\sum\nolimits_{i,j\in \hat{\mathcal{N}}_{k},i\leq j }A_{ij}}{\hat{n}_{k}(\hat{n}_{k}-1)/2},  &      & {k=l},   \\
\frac{\sum\nolimits_{i\in \hat{\mathcal{N}}_{k},j\in \hat{\mathcal{N}}_{l}}A_{ij}}{\hat{n}_{k}\hat{n}_{l}}, & & {k     \neq     l}. \\
\end{array} \right.
$$
Through plug in $
\hat{B}$ we get an new centered and rescaled $\widetilde{A}{'}$: 
$$
\widetilde{A}^{'}_{ij}=\left\{
\begin{array}{rcl}
\frac{A_{ij}-\hat{P}_{ij}}{\sqrt{n\hat{P}_{ij}(1-\hat{P}_{ij})}}     &      & {i     \neq     j},\\
	0    &      & {i=j}.\\
\end{array} \right.
$$
		
To establish the theoretical results, we give the following  assumptions:
\begin{assumption}\label{as1}
There exists a constant $c_0 > 0$ such that $\min\limits_{1 \leq k \leq K} n_k\geq \frac{c_0n}{K}$ for all $n$.
\end{assumption}
\textcolor{blue}{\begin{assumption}\label{as2}
	The entries of $B$ are bounded away from 0 and 1, and $B$ has no identical rows.
\end{assumption}}
\begin{thm} \label{thm2}
Suppose that Assumption \ref{as1} and Assumption \ref{as2} are satisfied, then under the null hypothesis $H_0:K=K_{0}$, if $K_0 = o({\sqrt{n}})$, the estimated test statistic  
$\hat{T} = \frac{1}{\sqrt{6}}trace(\widetilde{A}^{'3}) \stackrel{d}{\longrightarrow} N(0,1).$
\end{thm}
		
We defer the proof to section \ref{pf}. Given Theorem \ref{thm2}, for the testing problem
\ref{eq1}, we reject $H_0$ when $|\hat T| \geq t_{1-\alpha/2}$, where $t_{\alpha}$ is the $\alpha$-th quantile of $N(0,1)$.
		
\subsection{Hypothesis testing algorithm}
In this section, we put forward a sequential hypothesis testing algorithm to find the most proper number of communities. We can see from Theorem \ref{thm2} that our test statistic $\hat{T}$ converges to $N(0,1)$. By using this result, we 
propose the following algorithm \ref{alg1} to determine the number of communities. We use the spectral clustering that was introduced in Von Luxburg \cite{vonluxburgTutorialSpectralClustering2007} in the third step of algorithm \ref{alg1}. Note that the community detection method is irrelevant to our algorithm. It is feasible for one to use any other community detection method.
\begin{algorithm}[h]
	\caption{\textcolor{blue}{Hypothesis Testing Algorithm}}
	\label{alg1}
	
 \textcolor{blue}{\textbf{Function} \textsc{HypothesisTest}$(A)$}	
	\begin{algorithmic}[1]
		\State \textcolor{blue}{Initialize $K_0 = 1$}
		\While{\textcolor{blue}{$|\hat{T}(K_0)| \geq t_{1-\alpha/2}$}}
		\State \textcolor{blue}{\textbf{Function} \textsc{CommunityDetect}$(A, K_0)$}
		\State \textcolor{blue}{Compute $\hat{T}(K_0) = \frac{1}{\sqrt{6}}\mathrm{trace}(\widetilde{A}^{'3})$}
		\State \textcolor{blue}{$K_0 = K_0 + 1$}
		\EndWhile
		\State \textcolor{blue}{$\hat{K} = K_0$}
	\end{algorithmic}
\end{algorithm}
		
\section{Numerical experiments} \label{numexp}
Note that in the following simulations spectral clustering algorithm \cite{vonluxburgTutorialSpectralClustering2007} is used to derive the estimated label $\hat{g}$.
		
\subsection{The null distribution}
We conduct numerical experiments to show the performance of $\hat{T}$ under the null hypothesis in finite sample size in this section. It verifies the results in  Theorem \ref{thm1} and Theorem \ref{thm2}.
We follow the same set-ups in Lei \cite{leiGoodnessoffitTestStochastic2016}: there are two equal-sized
communities, with $B_{11} = B_{22 }= 0.7$ and $B_{12 }= B_{21 }= 0.3$,
as to sample sizes we consider a small network with $n = 50$.
		
\begin{figure}[h]
\centering
\begin{minipage}[t]{0.48\textwidth}
\centering
\includegraphics[height= 5.6cm,width=7.3cm]{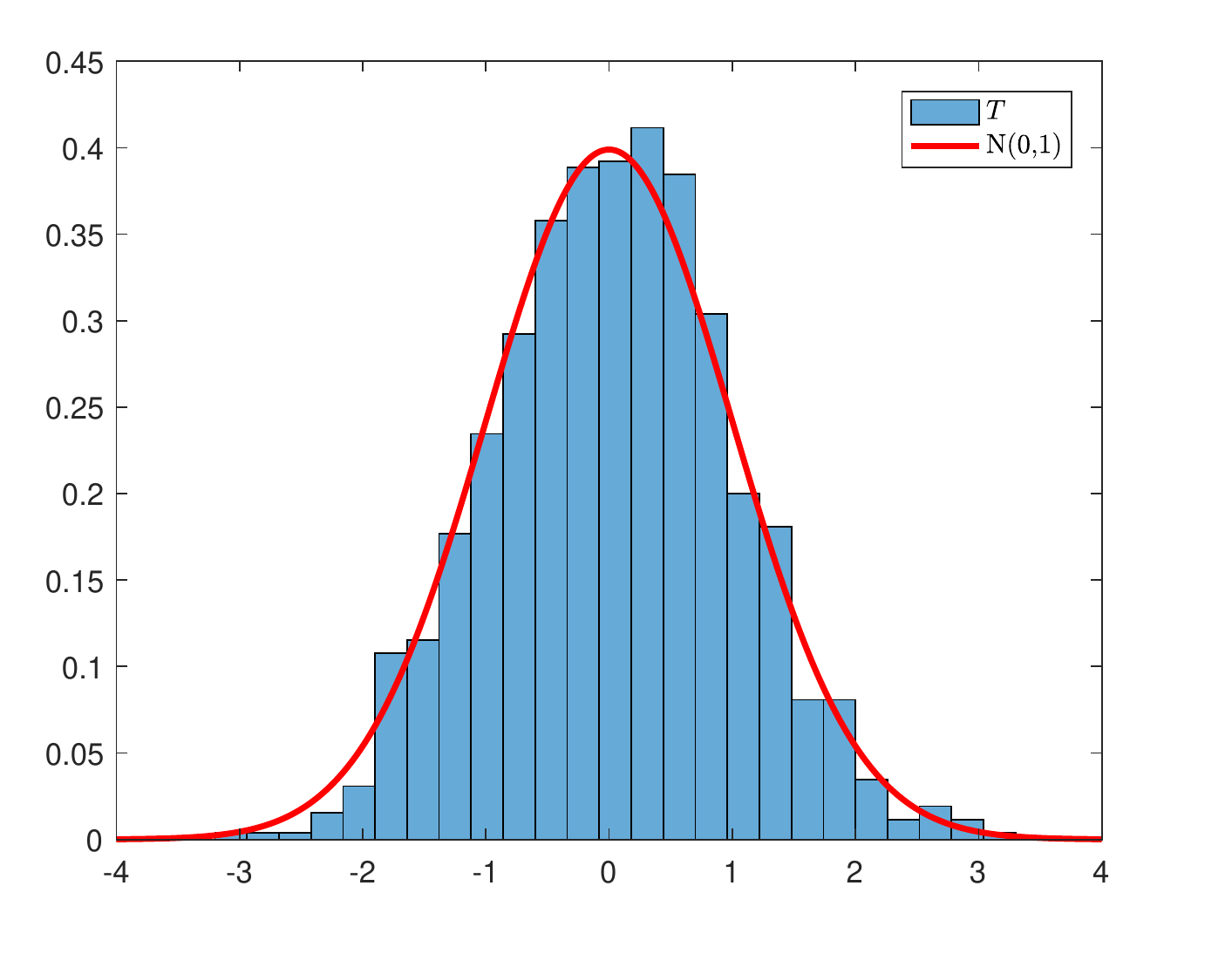}
				
\end{minipage}
\begin{minipage}[t]{0.48\textwidth}
\centering
\includegraphics[height= 5.5cm,width=7.3cm]{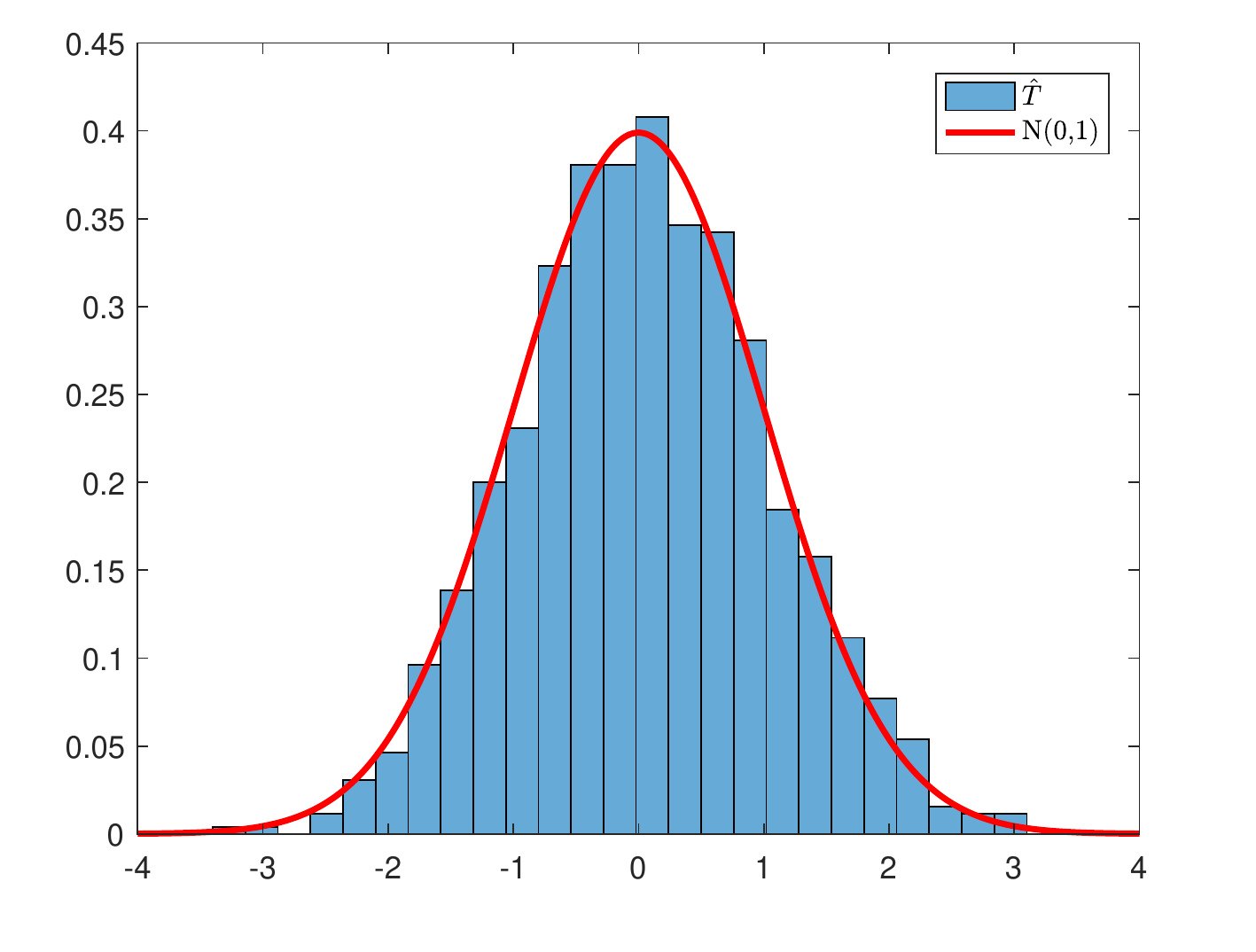}
				
\end{minipage}
\caption{ Histogram plots of the test statistic and pdf of $N(0,1)$. Left: $P$ is known; Right: $P$ is unknown.}\label{fig1}
\end{figure}

We present the histogram plots of our test statistic $\hat{T}$ from 1000 independent realizations under $H_0$ in Figure \ref{fig1}. The probability density function of $N(0,1)$ (red line) is  plotted as the comparison. It can visually confirm the results in Theorem \ref{thm1} and Theorem \ref{thm2}.

\subsection{The size and power of $\hat{T}$}
In this simulation, we will examine the size of our proposed test statistic under $H_0: K = K_0$, and the power under $H_1: K = K_0+1$. 
The edge probability between communities $k$ and $l$ is 0.2+0.4$\times I\{k = l\}$. Each entry of $g$ is sampled independently from the set $\{1, 2, \cdots, K\}$ with an equal probability. The network size $n = 1000$, after repeatedly performing the proposed hypothesis test 200 times, the reject rate at nominal level 0.05 are shown in Table \ref{tab1}. One can find that the size of our proposed test statistic is close to 0.05, and our test statistic is powerful from this table.
		
\begin{table}[h]
\centering
\setlength{\tabcolsep}{7mm}
\caption{\label{tab1}The reject rate at nominal level 0.05 over 200 trials.}
\renewcommand\arraystretch{1.5}
\begin{tabular}{lccl}
\toprule
$K_0$ & $K=K_0$ &$K=K_0+1$  \\
\midrule
2 & 0.04 & 1 \\
3 & 0.06 & 1 \\
4 & 0.05 & 1 \\
\bottomrule
\end{tabular}
\end{table}

Next, we will compare the performance of our test statistic $\hat{T}$ with that of Lei \cite{leiGoodnessoffitTestStochastic2016} by using their test statistic, which we refer to as $TW_1$. When the size of the network is small, $TW_1$ will not have good performance, so we will employ the bootstrap correction procedure as in Lei's paper. The true community number $K$ and the hypothesis one $K_0$ will vary from 2 to 5, and the size of the network is fixed at $n=1000$. Each entry of $g$ is sampled independently from the set $\{1, 2, \cdots, K\}$ with an equal probability. $B$ is the edge probabilities matrix with elements  $B_{kl}=0.3+4\times I\{k = l\}$, for $1\leq k,l\leq K$.
		
Under 100 independent replications, the reject rate at the nominal significance level of 0.05 can be seen in Table \ref{tab2}. One can see that our proposed test statistic $T$ has comparable performance with $TW_1$ from Table \ref{tab2}.
		
\begin{table}[h]
\caption{\label{tab2}The rejection rate at nominal level 0.05 over 100 trials.}
\centering
\setlength{\tabcolsep}{4mm}
\renewcommand\arraystretch{1.5}
\begin{tabular}{c|cccc|cccc}
\hline
& \multicolumn{4}{c|}{$\hat{T}$} & \multicolumn{4}{c}{$TW_1$ } \\
\hline
$K$      & 2 & 3 & 4 & 5                  & 2 & 3 & 4 & 5   \\
\hline
				
$K_0=2$     &$\bf{0.05}$&1.00&1.00&1.00
&$\bf{0.04}$&1.00&1.00&1.00\\
$K_0=3$     &*&$\bf{0.05}$&1.00&1.00
&*&$\bf{0.07}$&1.00&1.00\\
$K_0=4$     &*&*&$\bf{0.02}$&1.00
&*&*&$\bf{0.03}$&1.00\\
$K_0=5$     &*&*&*&$\bf{0.05}$&
*&*&*&$\bf{0.06}$\\
\hline
\end{tabular}
\end{table}
		
\subsection{Estimating $K$ by algorithm \ref{alg1}}
In the third simulation, we examine the performance of algorithm \ref{alg1}. The edge probabilities between communities $k$ and $l$ are $B_{kl}=r(3+4\times I\{k = l\})$, where $r$ controls the sparsity of the network. We consider $r \in \{0.01, 0.05, 0.1 \}$, and values of $K$ vary from 2 to 5. Each entry of $g$ is sampled independently from the set $\{1, 2, \cdots, K\}$ with an equal probability. Next, we generate 200 independent adjacency matrices $A$ with a network size of $n = 1000$ and a matrix $B$ for both $K$ and $r$. The proportion of correct estimates is listed in Table \ref{tab3}. One can see that our proposed algorithm \ref{alg1} has good performance for $K = 2, 3$ at all sparsity levels from Table \ref{tab3}.  Algorithm \ref{alg1} requires a more dense network to have good performance when $K$ gets larger, $K = 4, 5$. \textcolor{blue}{As a comparison, we also display the result by Lei's \cite{leiGoodnessoffitTestStochastic2016} method in Table \ref{tab3}. It can be seen that our method has comparable performance with his.}

\begin{table}[h]
	\caption{\label{tab3}The accuracy rate of estimating $K$ over 200 trials varies depending on the degree of sparsity, as indicated by the index $r$.}
	\centering
	\color{blue}
	\setlength{\tabcolsep}{4mm}
	\renewcommand\arraystretch{1.5}
	\begin{tabular}{c|ccc|ccc}
		\hline
		& \multicolumn{3}{c|}{$\hat{T}$} & \multicolumn{3}{c}{$TW_1$ } \\
		\hline
		$r$      & 0.01 & 0.05 & 0.1                  & 0.01 & 0.05 & 0.1   \\
		\hline
		
		$K=2$     &1&1&1
		& 1 & 1 & 1 \\
		$K=3$     &1&1&0.99
		& 1 & 0.99 &0.99\\
		$K=4$     & 0.26 & 0.95 &0.99
		& 0.24 & 0.90 &0.92\\
		$K=5$     & 0.00 & 0.87 &0.93
		& 0.01 & 0.74 &0.93\\
		\hline
	\end{tabular}
\end{table}

\section{Real Data Example} \label{realdata}
\subsection{The dolphin network}
\begin{figure}[h]
\centering
\begin{subfigure}[b]{0.4\textwidth}
\centering
\includegraphics[height= 4cm,width=\textwidth]{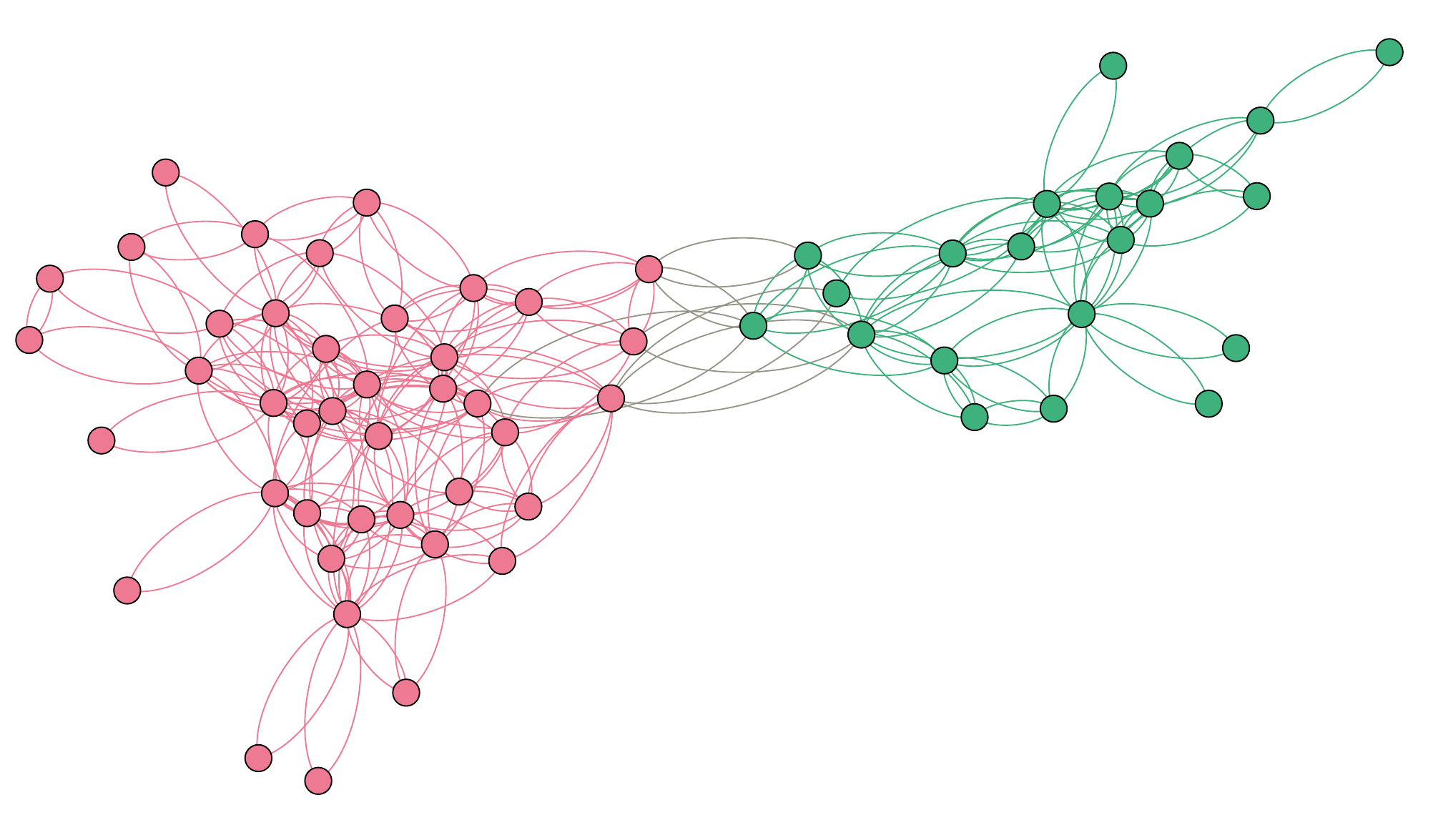}
\caption{Two communities}
\label{}
\end{subfigure}
\hfill
\begin{subfigure}[b]{0.4\textwidth}
\centering
\includegraphics[height= 4cm,width=\textwidth]{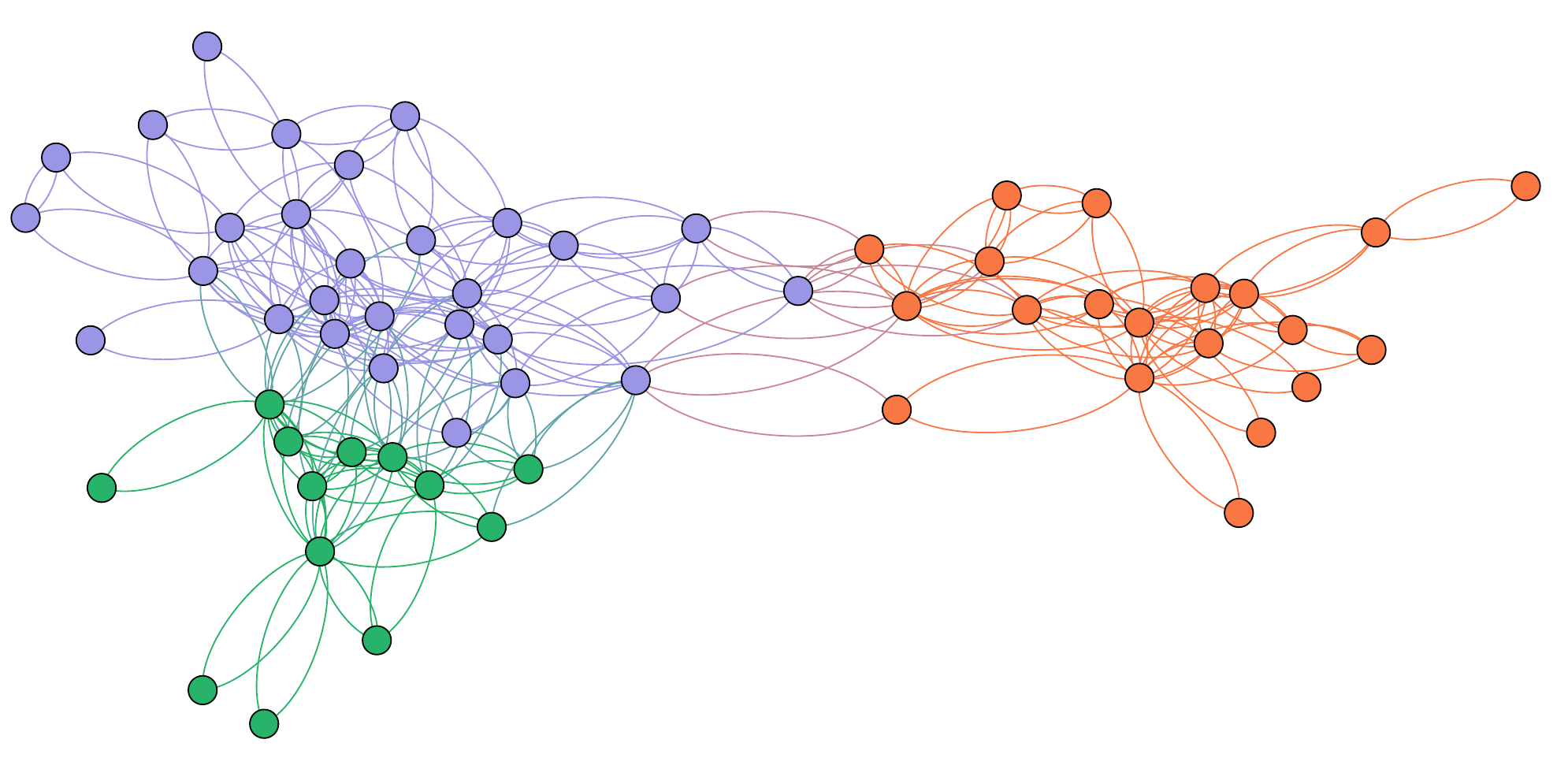}
\caption{Three communities}
\label{}
\end{subfigure}
\hfill
\begin{subfigure}[b]{0.4\textwidth}
\centering
\includegraphics[height= 4cm, width=\textwidth]{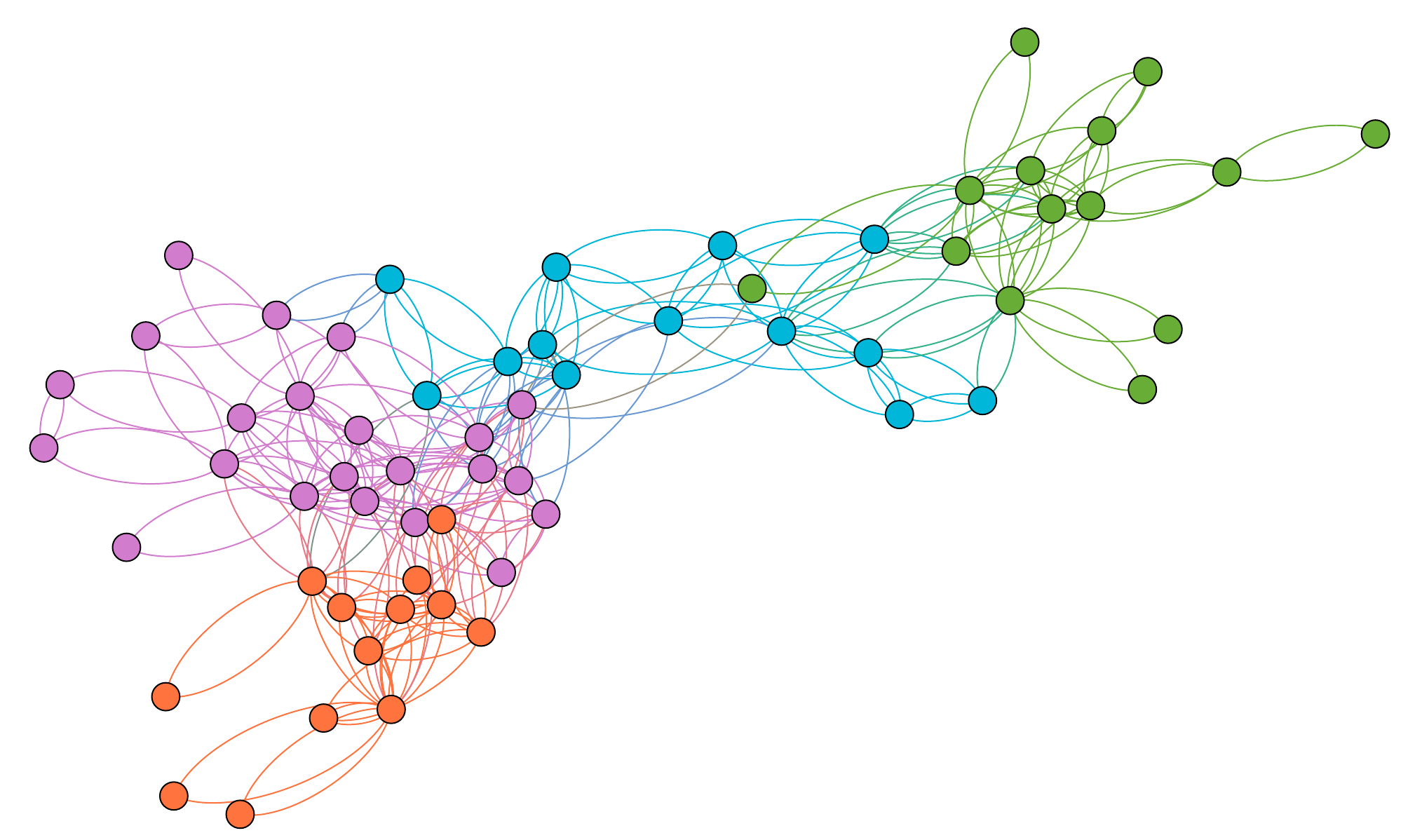}
\caption{Four communities}
\label{}
\end{subfigure}
\caption{Community detection results for varying community numbers.}
\label{comde}
\end{figure}

In this subsection we turn our attention to a popularly studied dolphin network \cite{lusseauBottlenoseDolphinCommunity2003}. 
This dolphin network is comprised of 62 nodes and 159 edges, with the nodes representing individual dolphins. From 1995 to 2001, adult members of each dolphin school encountered in the fjord were photographed based on natural markings on their dorsal fins. This data was used to analyze the frequency of dolphin pairs seen together, which ultimately led to the creation of a social network comprising 62 dolphins and 159 undirected edges representing their preferred companionship. Initially, this network was thought to be divisible into two distinct groups. Now it was argued in  \cite{liuDiscoveringCommunitiesComplex2016} that $K = 4$ is also reasonable. The p-values corresponding to $K\in\{2,3,4\}$ are listed in Table \ref{tab4}. In Figure \ref{comde} we present the community detection results that are obtained by using different $K$.
		
\begin{table}[h]
\centering
\setlength{\tabcolsep}{7mm}
\renewcommand\arraystretch{1.5}
\caption{\label{tab4} p-values correspond to $K\in\{2,3,4\}$. }
\begin{tabular}{lcccl}
\toprule
$K$ & 2&3  & 4 \\
\midrule
p-value & 0.0296 & 0.0208 & 0.0433  \\
\bottomrule
\end{tabular}
\end{table}

\subsection{The political blog data}
\begin{figure}[h]
\centering         
\includegraphics[height= 6cm,width=10cm]{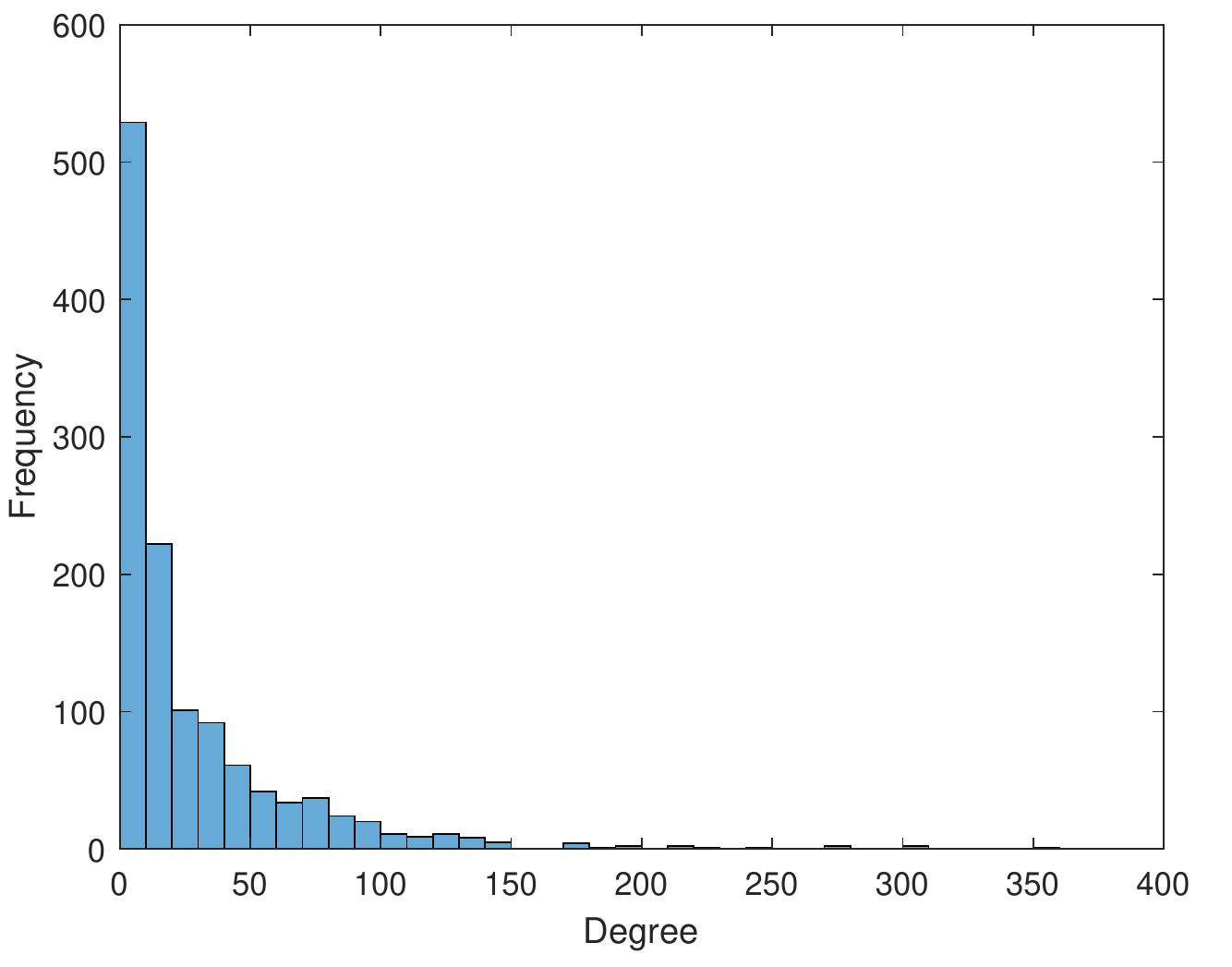} 
\caption{Degree histograms for the political blog network.}
\label{figDegree}
\end{figure}

In this subsection, we consider a well-known network collected by Adamic \cite{adamicPoliticalBlogosphere20042005}. In this dataset, nodes represent political blogs and edges represent hyperlinks between  blogs. 
These hyperlinks were recorded just before the 2004 US presidential election. As it is commonly done in the literature \cite{wangLikelihoodbasedModelSelection2017,zhaoConsistencyCommunityDetection2012}, our attention will be directed solely towards the largest connected component, which consists of 1222 nodes in this paper.
It can be divided the political blog network into two communities: the liberal and conservative, but there exist high degree nodes (see Figure \ref{figDegree}), so it is more proper to fit this network by the DCBM \cite{leiGoodnessoffitTestStochastic2016,karrerStochasticBlockmodelsCommunity2011}.  Under the two communities' assumption, the derived  $p$-value is 0. As result it is inappropriate to fit this network by the SBM with two communities. 
		
\section{Technical proofs}\label{pf}
In this section, we will give the proofs of Theorems \ref{thm1}-\ref{thm2}. We first introduce some definitions and recent progress in random matrix theory(RMT).
		
Denote $W$ be any symmetric matrix whose dimension is $n\times n$ and its eigenvalues are $\lambda_{i}$, $i=1,2,\cdots,n$. Next we define the empirical spectral distribution  of $W$ as follows:
$$
F_{n}(u)=\frac{1}{n}\sum_{i=1}^{n}I_{\{\lambda_{i}\leq u\}} ,
$$
where $I$ is the indicator function.
		
It has been proved in Bai et al. \cite{baiSpectralAnalysisLarge2010a} that $F_{n}(u)$ converges to $F(u)$ almost surely under some conditions, see the following Lemma \ref{lem1}, where $F(u)$ is the well known semicircular law and its probability density function is 
$$
F(du)=\frac{\sqrt{4-u^{2}}}{2\pi},\quad \forall u\in [-2,2].
$$
\begin{lem}(Theorem 2.9 in Bai et al. \cite{baiSpectralAnalysisLarge2010a}).\label{lem1}
Let $W_n = \frac{1}{\sqrt{n}} X_n$ be a symmetric matrix, we call it a Wigner matrix if the entries above or on the diagonal of $X_n$ are independent.
The subscript $n$ means that the entries of $X_n$ may be dependent on $n$, for simplicity we omit the subscript $n$. If for  $1\leq i,j \leq n$, we have $E(X_{i,j})=0$ and $Var(X_{i,j})=1$, given any positive constant $\eta$, the following condition is satisfied,
$$
(A.1)\quad\quad \lim\limits_{n\to\infty} \frac{1}{n^{2}}\sum_{i,j}E(|X_{i,j}|^{2})I_{\{|X_{i,j}|\geq \eta\sqrt{n}\}}=0.
$$
Then we have $F_{W_n}(u)$ converges to $F(u)$ almost surely.
\end{lem}
		
The domain of $F$ is the interval $[-2,2]$, actually we call $[-2,2]$ the support of $F$ in probability theory. We first define an open set $U$ such that $[-2,2]\subset U\subset R$, where $R $ is the set of real numbers. Next then the empirical process $G_n={G_n(f(u))}$ is given by:
$$(A.2)\quad\quad {G_n(f(u))}=n \int_{-\infty}^{\infty} f(u)[F_n(du)-F(du)]\ , \qquad \forall f\in \mathcal{A}, $$
where $\mathcal{A}$ represents the set of analytic functions $f:U\rightarrow R $.
		
Under following moment conditions(1)-(3), Wang et al. \cite{wangGeneralizationCLTLinear2021} proved the empirical process $G_n(f(u))$ converges weakly to $G(f(u))$, where $G(f(u))$ is Gaussian process, see Lemma \ref{lem2}.\\
\textcolor{blue}{(1) For  $1\leq i \leq n$, $E(|W_{i,i}|^{2})=\frac{\sigma^{2}}{n}$, and for $1\leq i < j\leq n$, we have $E(|W_{i,j}|^{2})=\frac{1}{n}.$\\
	(2) $\lim_{n \rightarrow \infty}\sum_{i,j}E|W_{i,j}|^{4}=L$.\\
	(3)The random variables $\sqrt{n}W_{ij}$ are bounded in any $L^p$ space ($p \geq 1$), uniformly in $n, i, j$.}

\begin{lem}(Theorem 2.1 in  Wang et al. \cite{wangGeneralizationCLTLinear2021})\label{lem2}.
If conditions (1)–(3) are satisfied, then we have that the spectral empirical process
$\{G_{n}(f), f\in \mathcal{A}\}$ converges to a Gaussian process $G := \left\{G(f) : f\in \mathcal{A}\right\}$ weakly
in finite dimension, where $\mathcal{A}$ represents the set of analytic functions.  The mean
function of $\{G(f), f\in \mathcal{A}\}$ is
$$
E[G(f)]=\frac{f(2)+f(-2)}{4}-\frac{1}{2}\tau_0(f)+(\sigma^{2}-2)\tau_2(f)+(L-3)\tau_4(f).
$$
The covariance of Gaussian process $\{G(f), f\in \mathcal{A}\}$ is
$$E((G(f)-E(G(f))(G(g)-E(G(g)))=cov(f,g)=\frac{1}{4\pi^2}\int_{-2}^{2} \int_{-2}^{2}f^{'}(x)g^{'}(y)V(x,y)\, dxdy
,$$
where $$
V(x,y)=(\sigma^{2}-2+\frac{L-3}{2} xy)\sqrt{(4-x^2)(4-y^2)}+2\log(\frac{4-xy+\sqrt{(4-x^2)(4-y^2)}}{4-xy-\sqrt{(4-x^2)(4-y^2)}}),
			$$	
			
$$
\tau_l(f)=\frac{1}{2\pi}\int_{-\infty}^{\infty} f(2cos\theta)cos(l\theta)\, d\theta.
$$
\end{lem}

\begin{pf}[Proof of Theorem \ref{thm1}]
In the first step, we prove the empirical spectral distribution of $\widetilde{A}$ converges to $F(u)$ almost surely. Denote $X=\sqrt{n}\widetilde{A}$. If the entries of matrix $X$ satisfy condition (A.1) then we have that the empirical spectral distribution of $\widetilde{A}$ converges to $F(u)$ almost surely since Lemma \ref{lem1}.
Given any positive constant $\eta$,

\begin{equation}
\begin{aligned}\nonumber
&\frac{1}{n^2}\sum_{i,j}E(|X_{i,j}|^{2})I_{\{|X_{i,j}|\geq \eta\sqrt{n}\}}
=\frac{1}{n^2}\sum_{i,j}I_{\{|X_{i,j}|\geq \eta\sqrt{n}\}}\\
&\leq\mathop{max}\limits_{ij}\left\{I_{\{|X_{i,j}|\geq \eta\sqrt{n}\}}\right\}
=\mathop{max}\limits_{i\neq j}\left\{I_{\{|X_{i,j}|\geq \eta\sqrt{n}\}}\right\}
\rightarrow 0.
\end{aligned}
\end{equation}
the last equality is due to $|X_{ii}|=0$, and $|X_{ij}|=\frac{A_{ij}-P_{ij}}{\sqrt{P_{ij}(1-P_{ij})}}$ is bounded.
			
Secondly, we prove that $\widetilde{A}$ satisfies conditions (1)-(3) of Lemma \ref{lem2}.
\textcolor{blue}{Since $E(|\widetilde{A}_{i,j}|^{2})=\frac{1}{n}-\delta_{ij}$ for  $ 1 \leq i,j \leq n,$ as 
a result condition (1) holds. Condition (2) holds because of $E|\widetilde{A}_{i,j}|^{4}=O(\frac{1}{n^2}).$
Last, we can verify that condition(3) holds.} 
			
Let $f(u)=u^{3}$, then its empirical process
\begin{equation}
\begin{aligned}\nonumber
&{G_n(u^{3})}=n \int_{-\infty}^{\infty} u^3F_n\, (du)-n \int_{-\infty}^{\infty} u^3F\, (du)\\
&~~\quad \quad=n \int_{-\infty}^{\infty} u^3F_n\, (du)
=\sum_{i=1}^{n}\lambda_{i}^{3}
=trace(\widetilde{A}^{3}).\\
\end{aligned}
\end{equation}
From Lemma \ref{lem2} we derive
$$ 
E(G (u^{3} )) = 0 , Var(G(u^{3})) = 6.
$$
This completes our proof.
\end{pf}
		
Next we will prove Theorem \ref{thm2} in matrix form. For simplicity, we introduce some notations used in the proof.
		
Given any matrices X and Y, we use $ X\circ Y$ to represent their Hadamard product whose entries are $(X\circ Y)_{ij}=(X)_{ij}(Y)_{ij}$,  $X\oslash Y$ represents the Hadamard division of any two matrices X and Y, whose entries are $(X\oslash Y)_{ij}=\frac{(X)_{ij}}{(Y)_{ij}},$ the Hadamard root of matrix $X$ is a matrix whose entries are given by  $(X^{\circ\frac{1}{2}})_{ij}= X_{ij}^{\frac{1}{2}}$. 
Note that $X$ and $Y$ should have the same dimension.
\begin{pf}[Proof of Theorem \ref{thm2}]
We first denote 
\begin{align}\label{eq4} 
\widetilde{A}_{1}=\widetilde{A}+(P-\hat{P})\oslash(nP\circ(J_n-P))^{\circ\frac{1}{2}},
\end{align}
\begin{align}\label{eq5} 
\widetilde{A_{1}^{'}}=(A-\hat{P}+\hat{P}\circ I_n)\oslash(nP\circ(J_n-P))^{\circ\frac{1}{2}}.
\end{align}
where  $\circ$ represents the Hadamard product, $\oslash$ represents the Hadamard division, ${\circ\frac{1}{2}}$  represents the Hadamard root, $J_n$ and $I_n$  have the same dimension $n\times n$, all entries in $J_n$ are one and $I_n$ is the traditional identify matrix.
			
From \ref{eq2} we have $\widetilde{A}=(A-P+P\circ I_{n})\oslash(nP\circ(J_n-P))^{\circ\frac{1}{2}}$,  then combining (\ref{eq4}) and (\ref{eq5}), we have
\begin{align}\label{eq6} 
\widetilde{A_{1}}=\widetilde{A_{1}^{'}}+((P-\hat{P})\circ I_n)\oslash(nP\circ(J_n-P))^{\circ\frac{1}{2}}.
\end{align}
			
From assumption \ref{as1}, we have $P_{ij}-\hat{P}_{ij}=O_p(\frac{k}{n})$, thus (\ref{eq6}) can be expressed as
\begin{align}\label{eq7} 
\widetilde{A_{1}}=\widetilde{A_{1}^{'}}+O_p(\frac{k}{n^\frac{3}{2}})I_n.	
\end{align}
Similarly, we have
\begin{align}\label{eq8} 
\widetilde{A}=\widetilde{A_{1}}+O_p(\frac{k}{{n^\frac{3}{2}}})J_n.
\end{align}
As a result, we have
$$
trace([\widetilde{A_{1}}]^{3}-[\widetilde{A_{1}^{'}}]^{3})=
O_p(\frac{k}{{n^\frac{3}{2}}})trace([\widetilde{A_{1}^{'}}]^{2})+O_p(\frac{k^{2}}{{n^{3}}})trace([\widetilde{A_{1}^{'}}])+O_p(\frac{k^{3}}{{n^\frac{9}{2}}})trace(I_{n}).
$$
First, we have
$trace(\widetilde{A^{'}_{1}})=0.$  Second, one can prove that
$$
trace([\widetilde{A_{1}^{'}}]^{2})=\sum_{i=1}^{n}\widetilde{[A_{1}^{'}}(i,i)]^{2}
+\sum_{i \neq j}[\widetilde{A_{1}^{'}}(i,j)]^{2}
=\sum_{i \neq j}[\widetilde{A_{1}^{'}}(i,j)]^{2}=O_p(n).
$$	
the last equality holds because of 
\begin{equation}
\begin{aligned}\nonumber
&\sum_{i \neq j}[\widetilde{A_{1}^{'}}(i,j)]^{2}=\sum_{i \neq j}\frac{(A_{ij}-P_{ij})^{2}}{nP_{ij}(1-P_{ij})}
+2\sum_{i \neq j}\frac{(A_{ij}-P_{ij})(P_{ij}-\hat{P}_{ij})}{nP_{ij}(1-P_{ij})}
+\sum_{i \neq j}\frac{(P_{ij}-\hat{P}_{ij})^2}{nP_{ij}(1-P_{ij})}\\
&~~~\quad \quad \quad \quad \quad=O_p(n)+O_p(k) +O_p(\frac{k^2}{n}).
\end{aligned}
\end{equation}	
Finally, we have
\begin{align}\label{eq9} 
trace([\widetilde{A_{1}}]^{3}-[\widetilde{A_{1}^{'}}]^{3})=
O_p(\frac{k}{{\sqrt n}}).
\end{align}
Similarly to equation (\ref{eq8}), one can derive
$$
trace([\widetilde{A_{1}}]^{3}-[\widetilde{A}]^{3})=
O_p(\frac{k}{{n^\frac{3}{2}}})trace([\widetilde{A}]^{2}J_{n})+O_p(\frac{k^{2}}{{n^{3}}})trace([\widetilde{A}J_{n}^{2}])+O_p(\frac{k^{3}}{{n^\frac{9}{2}}})trace(J_{n}^{3}).
$$
Note that
$$
trace([\widetilde{A}]^{2}J_{n})\leq \lambda_{max}([\widetilde{A}]^{2})n=O_P(n).
$$
$$
trace([\widetilde{A}]J_{n}^{2})\leq\lambda_{max}(\widetilde{A})n^{2}=O_P(n^{2}).
$$
As a result
\begin{align}\label{eq10} 
trace([\widetilde{A_{1}}]^{3}-[\widetilde{A}]^{3})=
O_p(\frac{k}{{\sqrt n}}).  
\end{align}
Recall that
\begin{align}\label{eq11} 
\widetilde{A}^{'}=(P\circ(J_n-P)\oslash(\hat{P}(J_n-\hat{P})))^{\circ\frac{1}{2}}\circ\widetilde{A_{1}^{'}},
\end{align} 
because of
\begin{align}\label{eq12} 
(P\circ(J_n-P)))^{\circ\frac{1}{2}}=(\hat{P}\circ(J_n-\hat{P})))^{\circ\frac{1}{2}}\circ(1+O_P(\frac{k}{n}))J_n.
\end{align}
we have
\begin{align}\label{eq13} 
\widetilde{A}^{'}=(1+O_P(\frac{k}{n}))\widetilde{A_{1}^{'}},
\end{align} 
As a result,
\begin{align}\label{eq14} 
trace([\widetilde{A^{'}}]^{3})=(1+O_P(\frac{k}{n}))^{3}trace([\widetilde{A^{'}_{1}}]^{3}).
\end{align}
Combing equation (\ref{eq9}) and (\ref{eq10}), we have
\begin{align}\label{eq15} 
trace([\widetilde{A^{'}}]^{3})=(1+O_P(\frac{k}{n}))^{3}(trace([\widetilde{A}]^{3})+O_p(\frac{k}{{\sqrt n}})).
\end{align}
This completes our proof.		
\end{pf}

%Next we consider the power of the test based on $\hat{\theta}$. The following
%theorem provides a lower bound of the growth rate of the test statistic $\hat{\theta}$
%under the alternative model $K >K_0$.	
		
%\begin{thm} 
%	Under the alternative hypothesis $H_1:K\textgreater K_{0}$, the estimated test statistic  
%	$\hat{\theta} = O_P(k\sqrt n).$
%\end{thm}
%\begin{pf}
%	Under the alternative hypothesis $P_{ij}-\hat{P}_{ij}=O_p(1)$, as result
%	$$\widetilde{A_{1}}=\widetilde{A_{1}^{'}}+O_p(\frac{k}{\sqrt n})I_n.$$
%	$$\widetilde{A}=\widetilde{A_{1}}+O_p(\frac{k}{{\sqrt n}})J_n.$$
%	$$
%	\widetilde{A}^{'}=(1+O_P(1))\widetilde{A_{1}^{'}}+O_P(1) C.
%	$$
%	Similar to Theorem 3.5, we have
%	\begin{equation}
%	\begin{aligned}\nonumber
%	&trace([\widetilde{A^{'}}]^{3})=(1+O_P(1))^{3}trace([\widetilde{A^{'}_{1}}]^{3})+(1+O_P(1))O_P(\sqrt n)\\
%	&\quad \quad \quad \quad \quad  =(1+O_P(1))^{3}(trace([\widetilde{A}]^{3})+O_p(k\sqrt n))+O_P(\sqrt n).
%	\end{aligned}
%	\end{equation}
%which completes our proof.
%\end{pf}
		
\section{Conclusion}\label{conclusion}
We propose a new method to test whether a network can be fitted by the SBM with $K_0$ communities in this paper. This test statistic is the linear spectral statistic about matrix $A$, which is the adjacency matrix of one network.  With some recent progress in random matrix theory, we derive the limit distribution of the linear spectral statistic in this paper, which is $N(0,1)$ under $H_0$. Simulations and real network datasets validate our test statistic has well performance. In many real-world networks, the nodes often exhibit degree heterogeneity.  This paper examines random networks using the framework of random matrix theory. Recently, there have been some results on block random matrices\cite{baoSpectralStatisticsSample2022,wangCentralLimitTheorem2021a}, we hope those results can help us extend the test statistic in this work to the more general degree corrected stochastic block models in the future.

\section*{Acknowledgments}
We would like to thank the Editor, Associate Editor, and the three referees for their constructive suggestions. Jiang Hu was partially supported by National Natural Science Foundation of China (Grant Nos. 12171078, 12292980, and 12292982), Fundamental Research Funds for the Central Universities No. 2412023YQ003 and National Key R \& D Program of China No. 2020YFA0714102.

		
		
		
		
%\begin{thebibliography}{00}
		

\bibliography{MyLibrary}

\begin{thebibliography}{29}
\expandafter\ifx\csname natexlab\endcsname\relax\def\natexlab#1{#1}\fi
\providecommand{\url}[1]{\texttt{#1}}
\providecommand{\href}[2]{#2}
\providecommand{\path}[1]{#1}
\providecommand{\DOIprefix}{doi:}
\providecommand{\ArXivprefix}{arXiv:}
\providecommand{\URLprefix}{URL: }
\providecommand{\Pubmedprefix}{pmid:}
\providecommand{\doi}[1]{\href{http://dx.doi.org/#1}{\path{#1}}}
\providecommand{\Pubmed}[1]{\href{pmid:#1}{\path{#1}}}
\providecommand{\bibinfo}[2]{#2}
\ifx\xfnm\relax \def\xfnm[#1]{\unskip,\space#1}\fi
%Type = Article
\bibitem[{Abbe(2018)}]{abbeCommunityDetectionStochastic2018}
\bibinfo{author}{Abbe, E.}, \bibinfo{year}{2018}.
\newblock \bibinfo{title}{Community {{Detection}} and {{Stochastic Block
  Models}}: {{Recent Developments}}}.
\newblock \bibinfo{journal}{Journal of Machine Learning Research}
  \bibinfo{volume}{18}, \bibinfo{pages}{1--86}.
%Type = Inproceedings
\bibitem[{Adamic and Glance(2005)}]{adamicPoliticalBlogosphere20042005}
\bibinfo{author}{Adamic, L.A.}, \bibinfo{author}{Glance, N.},
  \bibinfo{year}{2005}.
\newblock \bibinfo{title}{The political blogosphere and the 2004 {{U}}.{{S}}.
  election: Divided they blog}, in: \bibinfo{booktitle}{Proceedings of the 3rd
  International Workshop on {{Link}} Discovery},
  \bibinfo{publisher}{{Association for Computing Machinery}}. pp.
  \bibinfo{pages}{36--43}.
%Type = Article
\bibitem[{Amini et~al.(2013)Amini, Chen, Bickel and
  Levina}]{aminiPseudolikelihoodMethodsCommunity2013}
\bibinfo{author}{Amini, A.A.}, \bibinfo{author}{Chen, A.},
  \bibinfo{author}{Bickel, P.J.}, \bibinfo{author}{Levina, E.},
  \bibinfo{year}{2013}.
\newblock \bibinfo{title}{Pseudo-likelihood methods for community detection in
  large sparse networks}.
\newblock \bibinfo{journal}{The Annals of Statistics} \bibinfo{volume}{41},
  \bibinfo{pages}{2097--2122}.
%Type = Book
\bibitem[{Bai and Silverstein(2010)}]{baiSpectralAnalysisLarge2010a}
\bibinfo{author}{Bai, Z.}, \bibinfo{author}{Silverstein, J.W.},
  \bibinfo{year}{2010}.
\newblock \bibinfo{title}{Spectral Analysis of Large Dimensional Random
  Matrices}.
\newblock Springer {{Series}} in {{Statistics}}. \bibinfo{edition}{second} ed.,
  \bibinfo{publisher}{{Springer, New York}}.
%Type = Misc
\bibitem[{Bao et~al.(2022)Bao, Hu, Xu and
  Zhang}]{baoSpectralStatisticsSample2022}
\bibinfo{author}{Bao, Z.}, \bibinfo{author}{Hu, J.}, \bibinfo{author}{Xu, X.},
  \bibinfo{author}{Zhang, X.}, \bibinfo{year}{2022}.
\newblock \bibinfo{title}{Spectral {{Statistics}} of {{Sample Block Correlation
  Matrices}}}.
\newblock \href{http://arxiv.org/abs/arXiv:2207.06107}{{\tt
  arXiv:arXiv:2207.06107}}.
%Type = Article
\bibitem[{Bickel et~al.(2013)Bickel, Choi, Chang and
  Zhang}]{bickelAsymptoticNormalityMaximum2013}
\bibinfo{author}{Bickel, P.}, \bibinfo{author}{Choi, D.},
  \bibinfo{author}{Chang, X.}, \bibinfo{author}{Zhang, H.},
  \bibinfo{year}{2013}.
\newblock \bibinfo{title}{Asymptotic normality of maximum likelihood and its
  variational approximation for stochastic blockmodels}.
\newblock \bibinfo{journal}{The Annals of Statistics} \bibinfo{volume}{41},
  \bibinfo{pages}{1922--1943}.
%Type = Article
\bibitem[{Bickel and Sarkar(2016)}]{bickelHypothesisTestingAutomated2016}
\bibinfo{author}{Bickel, P.J.}, \bibinfo{author}{Sarkar, P.},
  \bibinfo{year}{2016}.
\newblock \bibinfo{title}{Hypothesis testing for automated community detection
  in networks}.
\newblock \bibinfo{journal}{Journal of the Royal Statistical Society: Series B
  (Statistical Methodology)} \bibinfo{volume}{78}, \bibinfo{pages}{253--273}.
%Type = Article
\bibitem[{Dong et~al.(2020)Dong, Wang and
  Liu}]{dongSpectralBasedHypothesis2020}
\bibinfo{author}{Dong, Z.}, \bibinfo{author}{Wang, S.}, \bibinfo{author}{Liu,
  Q.}, \bibinfo{year}{2020}.
\newblock \bibinfo{title}{Spectral based hypothesis testing for community
  detection in complex networks}.
\newblock \bibinfo{journal}{Information Sciences} \bibinfo{volume}{512},
  \bibinfo{pages}{1360--1371}.
%Type = Article
\bibitem[{Holland et~al.(1983)Holland, Laskey and
  Leinhardt}]{hollandStochasticBlockmodelsFirst1983}
\bibinfo{author}{Holland, P.W.}, \bibinfo{author}{Laskey, K.B.},
  \bibinfo{author}{Leinhardt, S.}, \bibinfo{year}{1983}.
\newblock \bibinfo{title}{Stochastic blockmodels: {{First}} steps}.
\newblock \bibinfo{journal}{Social Networks} \bibinfo{volume}{5},
  \bibinfo{pages}{109--137}.
%Type = Article
\bibitem[{Jalan and Bandyopadhyay(2007)}]{jalanRandomMatrixAnalysis2007}
\bibinfo{author}{Jalan, S.}, \bibinfo{author}{Bandyopadhyay, J.N.},
  \bibinfo{year}{2007}.
\newblock \bibinfo{title}{Random matrix analysis of complex networks}.
\newblock \bibinfo{journal}{Physical Review E} \bibinfo{volume}{76},
  \bibinfo{pages}{046107}.
%Type = Article
\bibitem[{Ji and Jin(2016)}]{jiCoauthorshipCitationNetworks2016}
\bibinfo{author}{Ji, P.}, \bibinfo{author}{Jin, J.}, \bibinfo{year}{2016}.
\newblock \bibinfo{title}{Coauthorship and citation networks for
  statisticians}.
\newblock \bibinfo{journal}{The Annals of Applied Statistics}
  \bibinfo{volume}{10}, \bibinfo{pages}{1779--1812}.
%Type = Article
\bibitem[{Jin(2015)}]{jinFastCommunityDetection2015}
\bibinfo{author}{Jin, J.}, \bibinfo{year}{2015}.
\newblock \bibinfo{title}{Fast community detection by {{SCORE}}}.
\newblock \bibinfo{journal}{The Annals of Statistics} \bibinfo{volume}{43},
  \bibinfo{pages}{57--89}.
%Type = Book
\bibitem[{Johnson and
  Wichern(2007)}]{johnsonAppliedMultivariateStatistical2007}
\bibinfo{author}{Johnson, R.A.}, \bibinfo{author}{Wichern, D.W.},
  \bibinfo{year}{2007}.
\newblock \bibinfo{title}{Applied {{Multivariate Statistical Analysis}}}.
\newblock \bibinfo{edition}{6th edition} ed., \bibinfo{publisher}{{Pearson}},
  \bibinfo{address}{{Upper Saddle River, N.J}}.
%Type = Article
\bibitem[{Karrer and Newman(2011)}]{karrerStochasticBlockmodelsCommunity2011}
\bibinfo{author}{Karrer, B.}, \bibinfo{author}{Newman, M.E.J.},
  \bibinfo{year}{2011}.
\newblock \bibinfo{title}{Stochastic blockmodels and community structure in
  networks}.
\newblock \bibinfo{journal}{Physical Review E} \bibinfo{volume}{83},
  \bibinfo{pages}{016107}.
%Type = Article
\bibitem[{Lei(2016)}]{leiGoodnessoffitTestStochastic2016}
\bibinfo{author}{Lei, J.}, \bibinfo{year}{2016}.
\newblock \bibinfo{title}{A goodness-of-fit test for stochastic block models}.
\newblock \bibinfo{journal}{Annals of Statistics} \bibinfo{volume}{44},
  \bibinfo{pages}{401--424}.
%Type = Article
\bibitem[{Lei and Rinaldo(2015)}]{leiConsistencySpectralClustering2015}
\bibinfo{author}{Lei, J.}, \bibinfo{author}{Rinaldo, A.}, \bibinfo{year}{2015}.
\newblock \bibinfo{title}{Consistency of spectral clustering in stochastic
  block models}.
\newblock \bibinfo{journal}{Annals of Statistics} \bibinfo{volume}{43},
  \bibinfo{pages}{215--237}.
%Type = Article
\bibitem[{Liu et~al.(2016)Liu, Jiang, Pellegrini and
  Wang}]{liuDiscoveringCommunitiesComplex2016}
\bibinfo{author}{Liu, W.}, \bibinfo{author}{Jiang, X.},
  \bibinfo{author}{Pellegrini, M.}, \bibinfo{author}{Wang, X.},
  \bibinfo{year}{2016}.
\newblock \bibinfo{title}{Discovering communities in complex networks by edge
  label propagation}.
\newblock \bibinfo{journal}{Scientific Reports} \bibinfo{volume}{6},
  \bibinfo{pages}{22470}.
%Type = Article
\bibitem[{Lusseau et~al.(2003)Lusseau, Schneider, Boisseau, Haase, Slooten and
  Dawson}]{lusseauBottlenoseDolphinCommunity2003}
\bibinfo{author}{Lusseau, D.}, \bibinfo{author}{Schneider, K.},
  \bibinfo{author}{Boisseau, O.J.}, \bibinfo{author}{Haase, P.},
  \bibinfo{author}{Slooten, E.}, \bibinfo{author}{Dawson, S.M.},
  \bibinfo{year}{2003}.
\newblock \bibinfo{title}{The bottlenose dolphin community of {{Doubtful
  Sound}} features a large proportion of long-lasting associations}.
\newblock \bibinfo{journal}{Behavioral Ecology and Sociobiology}
  \bibinfo{volume}{54}, \bibinfo{pages}{396--405}.
%Type = Article
\bibitem[{Ma et~al.(2018)Ma, Wang and
  Yu}]{maSemisupervisedSpectralAlgorithms2018}
\bibinfo{author}{Ma, X.}, \bibinfo{author}{Wang, B.}, \bibinfo{author}{Yu, L.},
  \bibinfo{year}{2018}.
\newblock \bibinfo{title}{Semi-supervised spectral algorithms for community
  detection in complex networks based on equivalence of clustering methods}.
\newblock \bibinfo{journal}{Physica A: Statistical Mechanics and its
  Applications} \bibinfo{volume}{490}, \bibinfo{pages}{786--802}.
%Type = Book
\bibitem[{Newman(2018)}]{newmanNetworks2018}
\bibinfo{author}{Newman, M.}, \bibinfo{year}{2018}.
\newblock \bibinfo{title}{Networks}.
\newblock \bibinfo{publisher}{Oxford university press}.
%Type = Article
\bibitem[{Newman(2006)}]{newmanFindingCommunityStructure2006}
\bibinfo{author}{Newman, M.E.J.}, \bibinfo{year}{2006}.
\newblock \bibinfo{title}{Finding community structure in networks using the
  eigenvectors of matrices}.
\newblock \bibinfo{journal}{Physical Review E. Statistical, Nonlinear, and Soft
  Matter Physics} \bibinfo{volume}{74}, \bibinfo{pages}{036104, 19}.
%Type = Article
\bibitem[{Newman and Leicht(2007)}]{newmanMixtureModelsExploratory2007}
\bibinfo{author}{Newman, M.E.J.}, \bibinfo{author}{Leicht, E.A.},
  \bibinfo{year}{2007}.
\newblock \bibinfo{title}{Mixture models and exploratory analysis in networks}.
\newblock \bibinfo{journal}{Proceedings of the National Academy of Sciences}
  \bibinfo{volume}{104}, \bibinfo{pages}{9564}.
%Type = Article
\bibitem[{Pontes et~al.(2015)Pontes, Gir{\'a}ldez and
  {Aguilar-Ruiz}}]{pontesBiclusteringExpressionData2015}
\bibinfo{author}{Pontes, B.}, \bibinfo{author}{Gir{\'a}ldez, R.},
  \bibinfo{author}{{Aguilar-Ruiz}, J.S.}, \bibinfo{year}{2015}.
\newblock \bibinfo{title}{Biclustering on expression data: {{A}} review}.
\newblock \bibinfo{journal}{Journal of Biomedical Informatics}
  \bibinfo{volume}{57}, \bibinfo{pages}{163--180}.
%Type = Article
\bibitem[{{von Luxburg}(2007)}]{vonluxburgTutorialSpectralClustering2007}
\bibinfo{author}{{von Luxburg}, U.}, \bibinfo{year}{2007}.
\newblock \bibinfo{title}{A tutorial on spectral clustering}.
\newblock \bibinfo{journal}{Statistics and Computing} \bibinfo{volume}{17},
  \bibinfo{pages}{395--416}.
%Type = Article
\bibitem[{Wang and Bickel(2017)}]{wangLikelihoodbasedModelSelection2017}
\bibinfo{author}{Wang, Y.X.R.}, \bibinfo{author}{Bickel, P.J.},
  \bibinfo{year}{2017}.
\newblock \bibinfo{title}{Likelihood-based model selection for stochastic block
  models}.
\newblock \bibinfo{journal}{Annals of Statistics} \bibinfo{volume}{45},
  \bibinfo{pages}{500--528}.
%Type = Misc
\bibitem[{Wang and Yao(2021a)}]{wangCentralLimitTheorem2021a}
\bibinfo{author}{Wang, Z.}, \bibinfo{author}{Yao, J.}, \bibinfo{year}{2021}a.
\newblock \bibinfo{title}{Central limit theorem for linear spectral statistics
  of block-{{Wigner-type}} matrices}.
\newblock \href{http://arxiv.org/abs/arXiv:2110.12171}{{\tt
  arXiv:arXiv:2110.12171}}.
%Type = Article
\bibitem[{Wang and Yao(2021b)}]{wangGeneralizationCLTLinear2021}
\bibinfo{author}{Wang, Z.}, \bibinfo{author}{Yao, J.}, \bibinfo{year}{2021}b.
\newblock \bibinfo{title}{On a generalization of the {{CLT}} for linear
  eigenvalue statistics of {{Wigner}} matrices with inhomogeneous fourth
  moments}.
\newblock \bibinfo{journal}{Random Matrices: Theory and Applications} .
%Type = Article
\bibitem[{Westveld and Hoff(2011)}]{westveldMixedEffectsModel2011}
\bibinfo{author}{Westveld, A.H.}, \bibinfo{author}{Hoff, P.D.},
  \bibinfo{year}{2011}.
\newblock \bibinfo{title}{A mixed effects model for longitudinal relational and
  network data, with applications to international trade and conflict}.
\newblock \bibinfo{journal}{The Annals of Applied Statistics}
  \bibinfo{volume}{5}.
%Type = Article
\bibitem[{Zhao et~al.(2012)Zhao, Levina and
  Zhu}]{zhaoConsistencyCommunityDetection2012}
\bibinfo{author}{Zhao, Y.}, \bibinfo{author}{Levina, E.}, \bibinfo{author}{Zhu,
  J.}, \bibinfo{year}{2012}.
\newblock \bibinfo{title}{Consistency of community detection in networks under
  degree-corrected stochastic block models}.
\newblock \bibinfo{journal}{The Annals of Statistics} \bibinfo{volume}{40},
  \bibinfo{pages}{2266--2292}.

\end{thebibliography}
\bibliographystyle{elsarticle-harv}
		
%\end{thebibliography}
\end{document}